\documentclass{aastex62}

\newcommand{\ztfa}{high-gravity-BLAP-1}
\newcommand{\ztfb}{high-gravity-BLAP-2}
\newcommand{\ztfc}{high-gravity-BLAP-3}
\newcommand{\ztfd}{high-gravity-BLAP-4}
\newcommand{\kms}{\ensuremath{{\rm km}\,{\rm s}^{-1}}}

\newcommand{\msol}{M$_\odot \,$}
\newcommand{\rsol}{R$_\odot \,$}
\newcommand{\teff}{T$_{\rm eff}$}
\newcommand{\logg}{$\log(g)$}

\graphicspath{{./}{figures/}}

\received{January 1, 2018}
\revised{January 7, 2018}
\accepted{\today}
\submitjournal{ApJ}

\shorttitle{A new class of pulsating hot subdwarfs}
\shortauthors{Kupfer et al.}

\begin{document}

\title{A new class of large amplitude radial mode hot subdwarf pulsators}

\correspondingauthor{Thomas Kupfer}
\email{tkupfer@ucsb.edu}

\author[0000-0002-6540-1484]{Thomas Kupfer}
\affiliation{Kavli Institute for Theoretical Physics, University of California, Santa Barbara, CA 93106, USA}

\author[0000-0002-4791-6724]{Evan B. Bauer}
\affiliation{Department of Physics, University of California, Santa Barbara, CA 93106, USA}

\author[0000-0002-7226-836X]{Kevin B. Burdge}
\affiliation{Division of Physics, Mathematics and Astronomy, California Institute of Technology, Pasadena, CA 91125, USA}

\author[0000-0001-8018-5348]{Eric C. Bellm}
\affiliation{DIRAC Institute, Department of Astronomy, University of Washington, 3910 15th Avenue NE, Seattle, WA 98195, USA}

\author{Lars Bildsten}
\affiliation{Kavli Institute for Theoretical Physics, University of California, Santa Barbara, CA 93106, USA}
\affiliation{Department of Physics, University of California, Santa Barbara, CA 93106, USA}

\author{Jim Fuller}
\affiliation{Division of Physics, Mathematics and Astronomy, California Institute of Technology, Pasadena, CA 91125, USA}

\author[0000-0001-5941-2286]{JJ Hermes}
\affiliation{Department of Astronomy, Boston University, 725 Commonwealth Ave., Boston, MA 02215, USA}


\author[0000-0001-5390-8563]{Shrinivas R. Kulkarni}
\affiliation{Division of Physics, Mathematics and Astronomy, California Institute of Technology, Pasadena, CA 91125, USA}

\author{Thomas A. Prince}
\affiliation{Division of Physics, Mathematics and Astronomy, California Institute of Technology, Pasadena, CA 91125, USA}

\author[0000-0002-2626-2872]{Jan van~Roestel}
\affiliation{Division of Physics, Mathematics and Astronomy, California Institute of Technology, Pasadena, CA 91125, USA}

\author{Richard Dekany}
\affiliation{Caltech Optical Observatories, California Institute of Technology, Pasadena, CA 91125, USA}

\author[0000-0001-5060-8733]{Dmitry A. Duev}
\affiliation{Division of Physics, Mathematics and Astronomy, California Institute of Technology, Pasadena, CA 91125, USA}

\author{Michael Feeney}
\affiliation{Caltech Optical Observatories, California Institute of Technology, Pasadena, CA 91125, USA}

\author{Matteo Giomi} 
\affiliation{Humboldt Universit\"at zu Berlin, Newtonstra{\ss}e 15, 12489 Berlin, Germany}

\author[0000-0002-3168-0139]{Matthew J. Graham}
\affiliation{Division of Physics, Mathematics and Astronomy, California Institute of Technology, Pasadena, CA 91125, USA}

\author{Stephen Kaye}
\affiliation{Caltech Optical Observatories, California Institute of Technology, Pasadena, CA 91125, USA}

\author{Russ R. Laher}
\affiliation{IPAC, California Institute of Technology, 1200 E. California Blvd, Pasadena, CA 91125, USA}

\author{Frank J. Masci}
\affiliation{IPAC, California Institute of Technology, 1200 E. California Blvd, Pasadena, CA 91125, USA}

\author{Michael Porter}
\affiliation{Caltech Optical Observatories, California Institute of Technology, Pasadena, CA 91125, USA}

\author{Reed Riddle}
\affiliation{Caltech Optical Observatories, California Institute of Technology, Pasadena, CA 91125, USA}  

\author{David L. Shupe}
\affiliation{IPAC, California Institute of Technology, 1200 E. California Blvd, Pasadena, CA 91125, USA}

\author{Roger M. Smith}
\affiliation{Caltech Optical Observatories, California Institute of Technology, Pasadena, CA 91125, USA}

\author[0000-0001-6753-1488]{Maayane T. Soumagnac}
\affiliation{Benoziyo Center for Astrophysics, Weizmann Institute of Science, Rehovot, Israel}

\author[0000-0003-4373-7777]{Paula Szkody}
\affiliation{Department of Astronomy, University of Washington, Seattle, WA 98195, USA}

\author{Charlotte Ward}
\affiliation{Department of Astronomy, University of Maryland, College Park, MD 20742, USA}




\begin{abstract}
Using high-cadence observations from the Zwicky Transient Facility at low Galactic latitudes, we have discovered a new class of pulsating, hot, compact stars. We have found four candidates, exhibiting blue colors ($g-r\leq-0.1$\,mag), pulsation amplitudes of $>5\%$, and pulsation periods of $200 - 475$\,sec. Fourier transforms of the lightcurves show only one dominant frequency. Phase-resolved spectroscopy for three objects reveals significant radial velocity, \teff\,and \logg\, variations over the pulsation cycle, consistent with large amplitude radial oscillations. The mean \teff\, and \logg\, for these stars are consistent with hot subdwarf B (sdB) effective temperatures and surface gravities. We calculate evolutionary tracks using MESA and adiabatic pulsations using GYRE for low-mass helium-core pre-white dwarfs and low mass helium-burning stars. Comparison of low-order radial oscillation mode periods with the observed pulsation periods show better agreement with the pre-white dwarf models. Therefore, we suggest that these new pulsators and Blue Large-Amplitude Pulsators (BLAPs) could be members of the same class of pulsators, composed of young $\approx0.25-0.35$\,\msol\, helium-core pre-white dwarfs.
\end{abstract}



\section{Introduction} \label{sec:intro}
Subdwarf B stars (sdBs) are hot stars of spectral type B with luminosities below the main sequence. The formation mechanisms and evolution of sdBs is still debated, although most sdBs are thought to be helium (He)-burning stars with masses $\approx0.5$\,\msol\,and thin hydrogen envelopes (\citealt{heb86,heb09,heb16}). Amongst the sdB stars, two types of multi-periodic pulsators have been discovered, both with generally milli-mag amplitudes up to occasionally a few percent \citep{ost10}. On the hotter side (\teff$\gtrsim28,000$\,K) are the V361\,Hya stars which are pressure mode (p-mode) pulsators with typical periods of a few minutes \citep{kil97}. On the cooler side (\teff$\lesssim28,000$\,K) are the V1093\,Her stars which are gravity mode (g-mode) pulsators with periods of 45 min to 2 hours \citep{gre03}. Only a few sdB pulsators with a dominant radial mode are known, including Balloon 090100001 and CS\,1246, which show photometric amplitudes of up to $6\%$ \citep{ore04,bar10}.
Even before their discovery, the variability of sdBs was predicted to be caused by non-radial pulsation modes \citep{cha96} driven by the opacity bump due to partial ionization of iron \citep{cha97, fon03}. 

A new class of pulsating hot stars known as Blue Large-Amplitude Pulsators (BLAPs) was discovered by \cite{pie17}. BLAPs show similar effective temperatures (\teff) as the sdBs but lower surface gravities (\logg) and are an order of magnitude more luminous at $L\approx10^2-10^3$\,L$_\odot$ with pulsation periods between 20-40 minutes. Given their unusual location on the HR diagram, it has been proposed that BLAPs are low-mass ($M\approx0.3$\,\msol) helium-core pre-white dwarfs (pre-WDs). \citet{cor18} and \citet{rom18} showed that their pulsation periods can be explained by high-order non-radial g-modes or, in the case of the shortest periods, also by low-order radial modes. When including effects of radiative levitation of iron-group elements, \citet{byr18} found that the fundamental radial mode can be excited in a $0.31$\,\msol\, low-mass He-core pre-white dwarfs with \teff$\approx28,000$\,K, comparable to those of the BLAPs. Additionally, \citet{max13,max14} and \citet{gia16} reported the discovery of p-mode pulsations with periods in the range $\approx320-750$\,sec in extremely low-mass He-core pre-WDs (pre-ELMVs). \citet{jef13}, \citet{cor16} and \citet{ist17} found that the observed modes are consistent with theoretical predictions of p-mode pulsations in mixed-atmosphere He-core pre-WDs driven by the classical $\kappa$-mechanism operating in the partial helium ionization zones.

In this letter, we report the discovery of a new class of high-amplitude ($0.05 - 0.2$\,mag) pulsating sdB stars that show amplitudes similar to BLAPs, but spectral properties and pulsation periods similar to sdB p-mode pulsators (\teff$\approx32,000$\,K; pulsation periods $200 - 475$\,sec). We discovered four candidates of this new class; ZTF\,J071329.02$-$152125.2 (hereafter \ztfa), ZTF\,J184521.40$-$254437.5 (hereafter \ztfb),  ZTF\,J191306.79$-$120544.6 (hereafter \ztfc) and ZTF\,J182815.88$+$122530.5 (hereafter \ztfd). We suggest that these new pulsators (high-gravity-BLAPs) and BLAPs could be members of the same class of pulsators, composed of young $\approx 0.3 \, M_\odot$ He-core pre-WDs evolving through the sdB region of the \teff-\logg\,diagram. However, we show that we cannot exclude the inference that these new pulsators are low-mass He-shell burning stars with masses $\approx0.46$\,\msol, which have evolved off the extreme horizontal branch.

\section{Observations} \label{sec:observations}
As part of the Zwicky Transient Facility (ZTF), the Palomar 48-inch (P48) telescope images the sky every clear night.
Our four objects were discovered as part of a dedicated high-cadence survey at low Galactic latitudes with ZTF \citep{bel19,gra19}. During that dedicated survey we either observed one field or alternated between two adjacent fields continuously for $\approx$1.5-3\,hours on two to three consecutive nights in the ZTF-$r$ band \citep{bel19a}. Image processing of ZTF data is described in full detail in \citet{mas19}. The ZTF lightcurves of our candidates have $\approx$200-400 epochs observed over 2-3 nights in the ZTF-$r$ band. 

Additionally, 1.5\,hrs of high-cadence observations in the $g$ and $r$-bands using an exposure of 5\,sec were conducted for \ztfc, using the Palomar 200-inch telescope with the high-speed photometer CHIMERA \citep{har16}. For the same object, we also obtained $g$-band observations using the 2m telescopes of the Las Cumbres Observatories (LCO; \citealt{bro13}) at Siding Spring Observatory. We obtained a total of 130 exposures of $g$-band images, using an exposure time of 20\,sec. For \ztfa, we also obtained additional 220 $g$-band epochs with 8 sec exposures.


Phase-resolved spectroscopy for three candidates were obtained using the Keck\,I Telescope and the blue arm of the Low Resolution Imaging Spectrometer (LRIS; \citealt{mcc98}) using a low resolution mode ($R\sim1000$). To resolve the pulsation modes, we adopted a 20\,sec exposure time for \ztfa\, and a 45\,sec exposure time for \ztfb\,and \ztfc. We used $4\times4$ binning to reduce the large readout time to 27\,sec. We obtained a total of 75 spectra for \ztfa, 45 spectra for \ztfb\, and 26 spectra for \ztfc. Data reduction was performed with the Lpipe pipeline\footnote{http://www.astro.caltech.edu/\~dperley/programs/lpipe.html}\citep{per19}.
For \ztfd\, we obtained two spectra using the Double-Beam Spectrograph (DBSP; \citealt{oke82}) mounted on the Palomar 200-inch telescope. Each spectrum covered two pulsation cycles in the low resolution mode ($R\sim1500$). The data were reduced using a custom \texttt{PyRAF}-based pipeline \footnote{https://github.com/ebellm/pyraf-dbsp}\citep{bel16}. 


\begin{table}[t!]
\centering
	\caption{Photometric properties of the high-gravity-BLAPs}\label{tab:photresults}
  \begin{tabular}{lcccccc}
\tablewidth{0pt}
\hline
\hline
		 Object & RA (J2000) & Dec (J2000) &  $g^a$ & $g-r^a$ & $A_{\textrm {ZTF-r}}^b$ & $P$   \\ 
		        & ($^{\textrm h}:^{\textrm {min}}:^{\textrm {sec}}$)        & ($^\circ:^\prime:^{\prime\prime}$)        & (mag) & (mag) & (mmag)     & (sec)  \\
\hline
\ztfa & 07:13:29.02 & $-$15:21:25.2  &   16.53 &  $-$0.11  &  $53.9\pm2.5$    &  $200.20\pm0.02$  \\
\ztfb & 18:45:21.40 & $-$25:44:37.5  &   18.94 &	$-$0.09  &  $147.7\pm5.5$  &    $363.16\pm0.02$   \\
\ztfc & 19:13:06.79 & $-$12:05:44.6  &   17.57 &	$-$0.13  &  $129.1\pm3.3$  &    $438.83\pm0.01$  \\
\ztfd & 18:28:15.88 & +12:25:30.5  &   17.28 &	$-$0.14  &  $120.9\pm2.7$  &    $475.48\pm0.02$  \\
\hline
\hline
\end{tabular}
\begin{flushleft}
\centering
$^a$ taken from the Pan-STARRS release 1 (PS1) survey  \citep{cham16}\\
$^b$ amplitude in ZTF-$r$ from the ZTF lightcurves
\end{flushleft}
\end{table}

\begin{figure}
\begin{center}
\includegraphics[width=0.32\textwidth]{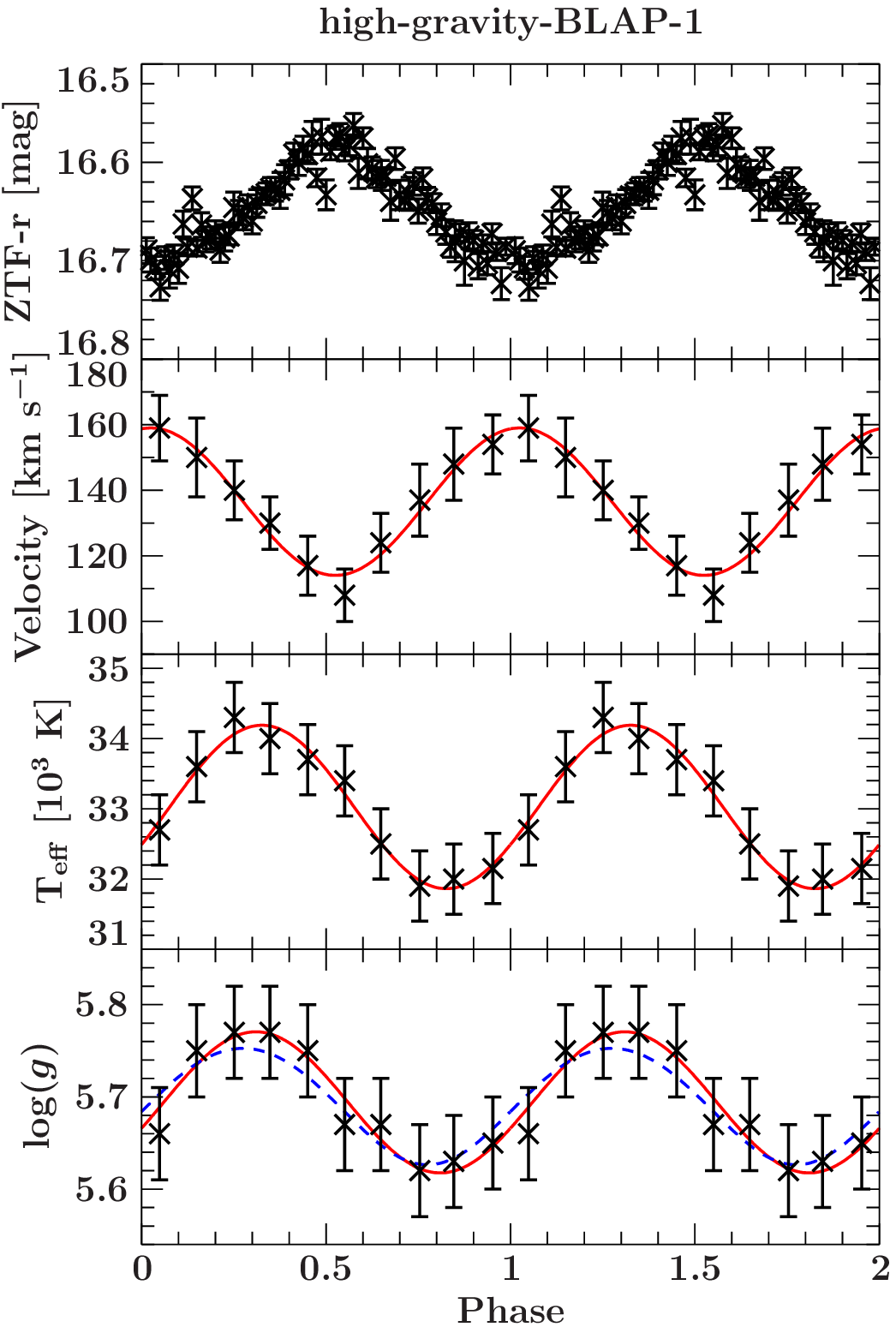}
\includegraphics[width=0.32\textwidth]{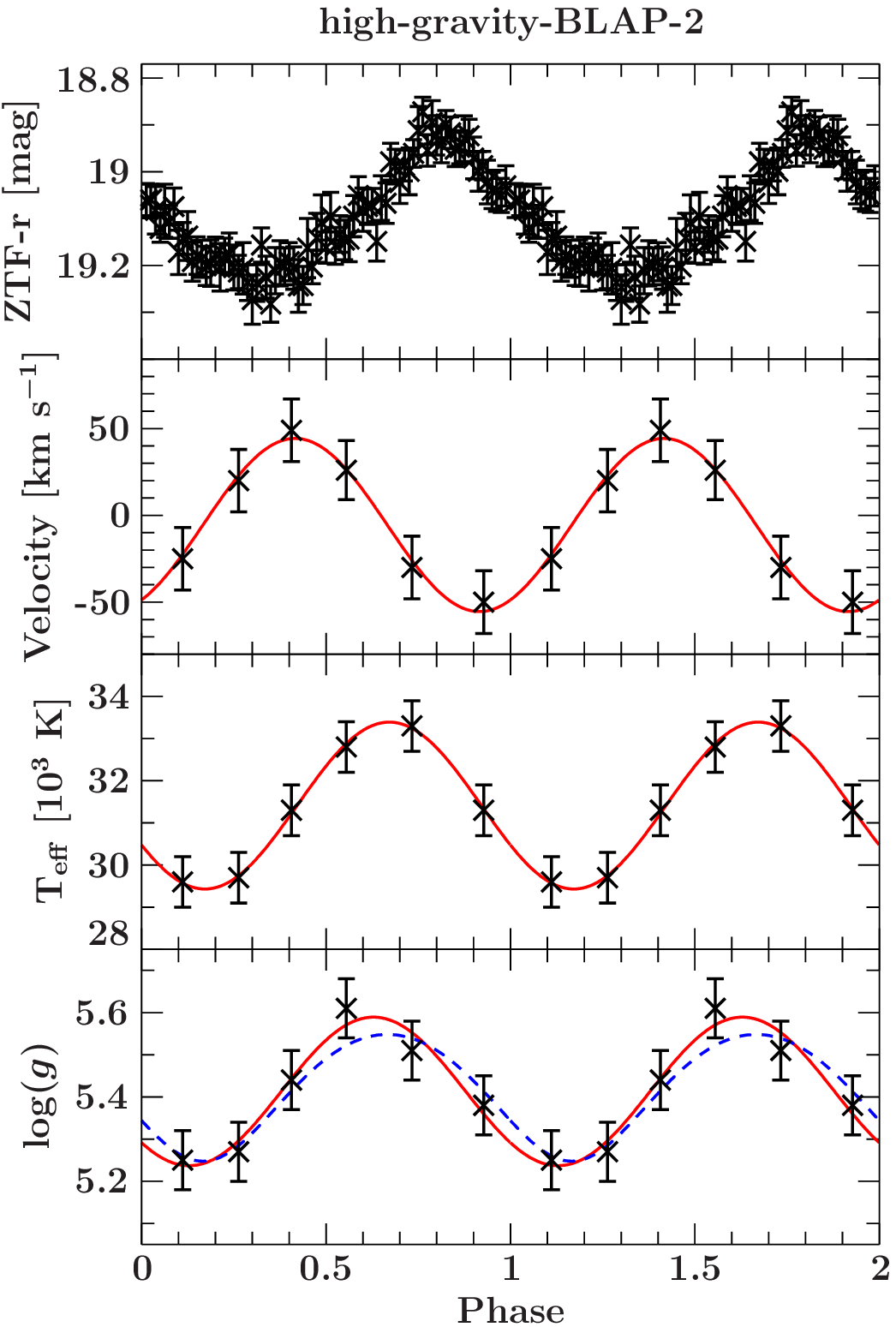}
\includegraphics[width=0.32\textwidth]{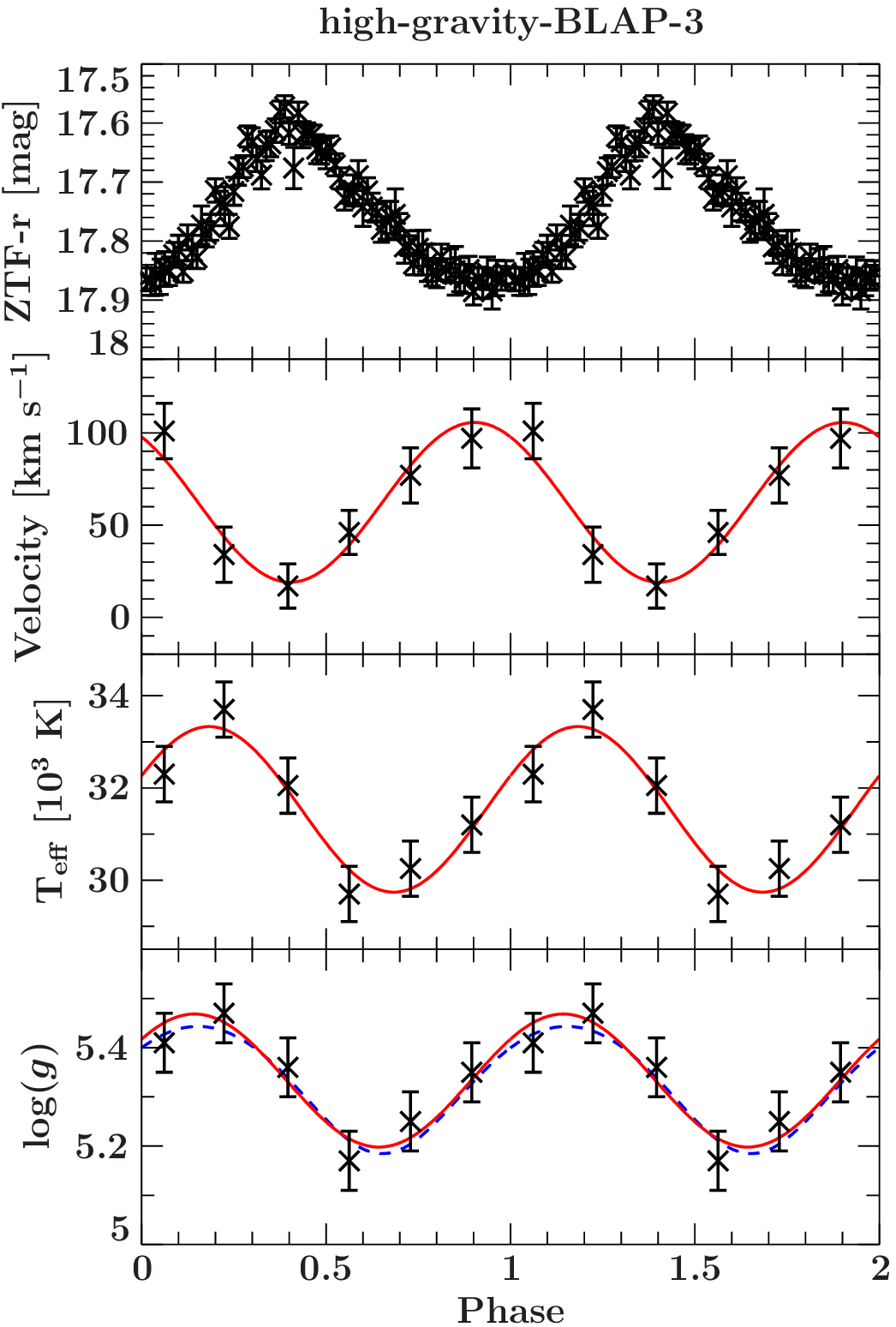}
\end{center}
\caption{{\bf Top:} Binned and phase-folded ZTF lightcurve against pulsation phase, {\bf Second from top:} Radial velocities against pulsation phase with a single harmonic fit (solid red curve), {\bf Third from top:} \teff\,against pulsation phase with a single harmonic fit (solid red curve), {\bf Bottom:} \logg\,against pulsation phase with the single harmonic fit (solid red curve) and the prediction from the fit to the radial velocities (dashed blue curve). The left panel shows the results for \ztfa, the middle panel for \ztfb, and the right panel for \ztfc. Two pulsation cycles are plotted for better visualization.}
\label{fig:pulsation}
\end{figure}

\section{Results}
\begin{figure}
\begin{center}
\includegraphics[width=0.5\textwidth]{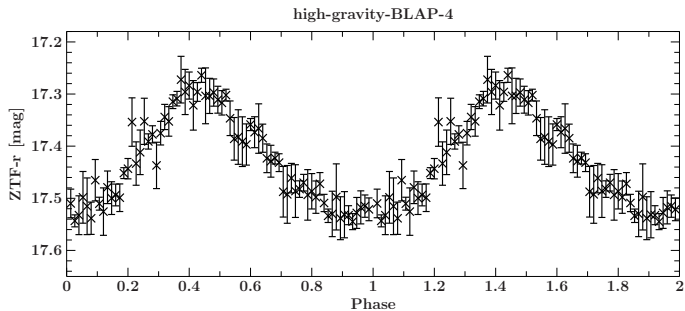}
\end{center}
\caption{Binned and phase-folded ZTF lightcurve against pulsation phase for \ztfd. Two pulsation cycles are plotted for better visualization. Phase-resolved spectroscopy was not obtained for this pulsator.}
\label{fig:pulsation1}
\end{figure}

All objects were initially discovered as periodic objects with blue colors and periods below $\approx$10\,min in a period search on the low Galactic latitude, high-cadence ZTF data, using the GPU implementation of the conditional entropy algorithm \citep{gra13}. All four objects stood out as short-period blue objects with photometric amplitudes larger than a few percent. The periods were refined using the \texttt{Gatspy} module for time series analysis that implements the Lomb-Scargle periodogram\footnote{http://dx.doi.org/10.5281/zenodo.14833}\citep{Lom76,sca82,van15} and allows different filter bands to be used in one fit. For \ztfb\, and \ztfd\, we only have the ZTF lightcurve whereas for \ztfc\, we combine the ZTF data with the LCO and Chimera data, and for \ztfa\,we combine ZTF data with LCO photometry. We find pulsation periods of $200.20\pm0.02$\,sec, $363.16\pm0.02$\,sec, $438.83\pm0.01$\,sec and $475.48\pm0.02$\,sec for \ztfa,  \ztfb, \ztfc\, and \ztfd\,respectively (see Table\,\ref{tab:photresults}). The photometric amplitude ($A_{\textrm {ZTF-r}}$) was derived from a Fourier analysis to the ZTF-$r$ band lightcurve using the \texttt{Period04} module\footnote{https://www.univie.ac.at/tops/Period04/}\citep{len05}. The phase-folded ZTF lightcurves are shown in Figs.\,\ref{fig:pulsation} and  \ref{fig:pulsation1}. 

The individual LRIS spectra of the three corresponding stars have relatively low signal-to-noise ratio (S/N) due to the short exposure time. To increase the S/N, the spectra were folded on the pulsation period into ten phase bins for \ztfa, and six phase-bins for \ztfb\,and \ztfc, respectively. We co-added individual spectra observed at the same pulsation phase. This increased the S/N per phase bin to $\approx30-40$. The \texttt{FITSB2} routine \citep{nap04a} was used to measure radial velocities. Lorentzian and Gaussian functions were fitted to the individual Balmer lines, excluding the H$_\epsilon$ line due to its blend with the interstellar Ca-H line. To obtain \teff, \logg, and the helium abundance ($\log[y]=\log[n_{\rm He}/n_{\rm H}]$), we fit a grid of metal-line-blanketed, local-thermodynamical equilibrium (LTE) atmospheres with solar metallicity \citep{heb00} and LTE models with enhanced metal-line blanketing and $10\times$solar metallicity \citep{oto06} to the individual phase-folded spectra. The spectral resolution of $\approx1000$ is insufficient to resolve metal lines; hence, we cannot measure the metallicty of these objects. We do not detect helium lines in the individual phase-folded spectra. Therefore, each spectrum was velocity corrected and co-added to increase the S/N and reveal weak helium features. We fit \teff, \logg, and helium abundance in the co-added spectra and kept the helium abundance fixed to the obtained value for each phase-bin, finding low helium abundances for all three stars with phase-resolved spectra; $\log(y)=-2.1\pm0.2$ for \ztfa, $\log(y)=-2.2\pm0.3$ for \ztfb\, and  $\log(y)=-2.0\pm0.2$ for \ztfc\, (see Table\,\ref{tab:specresults}). 

All three objects show significant radial velocity, \teff\,and \logg\,variations across the pulsation period. A single harmonic sine-curve was fitted to the radial velocities, effective temperatures and surface gravities as shown in Fig.\,\ref{fig:pulsation}. The results from the fit are presented in Table\,\ref{tab:specresults}. Additionally, the \logg\,amplitude was predicted from the measured velocities by calculating the time-derivative of the velocity fit following equation \ref{equ:logg} with $A_\textrm{RV}$ being the radial velocity amplitude, $\omega=\frac{2\pi}{P}$ and $\phi$ the pulsation phase:

\begin{equation}\label{equ:logg}
\log{(g)} = \langle \log{(g)} \rangle+A_\textrm{RV}\omega\cos{(\omega t+\phi)}     \,
\end{equation}

The predicted amplitude is consistent with the the observed \logg\,amplitude. We find radial velocity amplitudes of $22.5\pm3.0$\,\kms, $50.0\pm7.0$\,\kms\,and $43.3\pm6.5$\,\kms\,for \ztfa, \ztfb\, and \ztfc, respectively. 

Because our spectra for \ztfd\,are not phase-resolved, we can only measure the average $\langle$\teff$\rangle$, $\langle$\logg$\rangle$ and $\log(y)$, by co-adding the DBSP spectra to reach a S/N$\approx40-50$. Using the same LTE models, we find parameters consistent with an sdB star and a helium abundance $\log(y)=-2.4\pm0.4$. The average $\langle$\teff$\rangle$ is similar amongst all systems, whereas the average $\langle$\logg$\rangle$ decreases with increasing pulsation period. An overview of the spectroscopic results using the solar-metallicity as well as the $10\times$solar-metallicity models, is presented in Table\,\ref{tab:specresults}.

\begin{table}[t!]
\centering
	\caption{Spectroscopic properties of the high-gravity-BLAPs}\label{tab:specresults}
  \begin{tabular}{lccccccc}
\tablewidth{0pt}
\hline
\hline
		 Object  &  $A_\textrm{RV}$ &  $\langle$\teff$\rangle$ &  $\Delta$ \teff  & $\langle$\logg$\rangle$  &  $\Delta$\logg   &  $\Delta a$ & $\log(y)$  \\ 
		        &  (\kms)   &  (K) &  (K)  &   &  & (cm\,s$^{-2}$) &  \\
\hline
\ztfa\,(sol)  & $22.5\pm3.0$   &     $34\,000\pm500$      &  $1150\pm300$   &   $5.70\pm0.05$   &   $0.08\pm0.02$  & $85300\pm17500$  &  $-2.1\pm0.2$ \\ \smallskip
\ztfa\.($10\times$sol)  & $22.5\pm3.0$   &    $33\,400\pm500$   &   $1300\pm300$   &  $5.63\pm0.05$        &  $0.08\pm0.02$ & $72600\pm16800$ & $-2.1\pm0.2$  \\
\ztfb \,(sol)  & $50.0\pm7.0$ & $31\,400\pm600$  &  $2000\pm400$ & $5.41\pm0.06$ & $0.18\pm0.03$ & $99800\pm20800$ & $-2.2\pm0.3$ \\ \smallskip
\ztfb\,($10\times$sol)  & $50.0\pm7.0$ & $31\,600\pm600$  &  $2300\pm400$ & $5.36\pm0.06$ & $0.18\pm0.03$ & $87700\pm19700$ &  $-2.2\pm0.3$ \\
\ztfc\,(sol)  & $43.3 \pm6.5$ & $31\,600\pm600$  &  $1800\pm400$ &  $5.33\pm0.05$ & $0.14\pm0.03$ & $66800\pm18200$ & $-2.0\pm0.2$ \\ \smallskip
\ztfc\,($10\times$sol)  & $43.3 \pm6.5$ & $31\,800\pm600$  &  $2100\pm400$ &  $5.29\pm0.05$ & $0.13\pm0.03$ & $57700\pm17900$  & $-2.0\pm0.2$  \\
\ztfd\,(sol)  & - & $31\,700\pm500$  & - &  $5.31\pm0.05$ & -  &   - & $-2.4\pm0.4$ \\
\ztfd\,($10\times$sol)  & -  & $32\,000\pm500$  &  - &  $5.27\pm0.05$ & - &  -  & $-2.4\pm0.4$ \\
\hline
\hline
\end{tabular}
\end{table}

\section{Discussion}

\subsection{Absolute magnitude of \ztfa}
We estimate the absolute magnitude of the objects, using the distances from the Gaia DR2 parallaxes \citep{gai18}. For parallax measurements with fractional parallax errors $\sigma_{\varpi}/\varpi$ less than about $0.1-0.2$, the distance estimates are nearly independent of the choice of prior and can be calculated with $d=\varpi^{-1}$. For larger fractional errors, the estimated distance depends heavily on how well the prior reflects the true distribution of distances for the population of sources (e.g. \citealt{bai18} and references therein). As we have little knowledge of the true distribution of distances for the population of this class of pulsators, we present only the absolute magnitude for \ztfa, which is the only system with a fractional error $<0.2$. The observations reported in Gaia DR2 for \ztfa\,are $\varpi=0.7113\pm0.0854$\,arcsec and a Gaia G-band magnitude $G=16.54$. Using $d=\varpi^{-1}$, we find a distance $d=1.41^{+0.19}_{-0.15}$\,kpc which results in an absolute magnitude of $M_{\textrm G}=5.8\pm0.3$. 

There is substantial reddening towards \ztfa\, of around $\approx\,1-1.2$\,mag \citep{gre18}. Hence the absolute magnitude of \ztfa\,is around $M_{\textrm G}=6.8-7$. The typical absolute magnitude for hot subdwarfs are $7\lesssim M_{\textrm G} \lesssim 1$ \citep{gei19}. Therefore, we conclude that \ztfa\, falls on the faint end of the hot subdwarf regime.

\begin{figure}
\begin{center}
\includegraphics[width=0.99\textwidth]{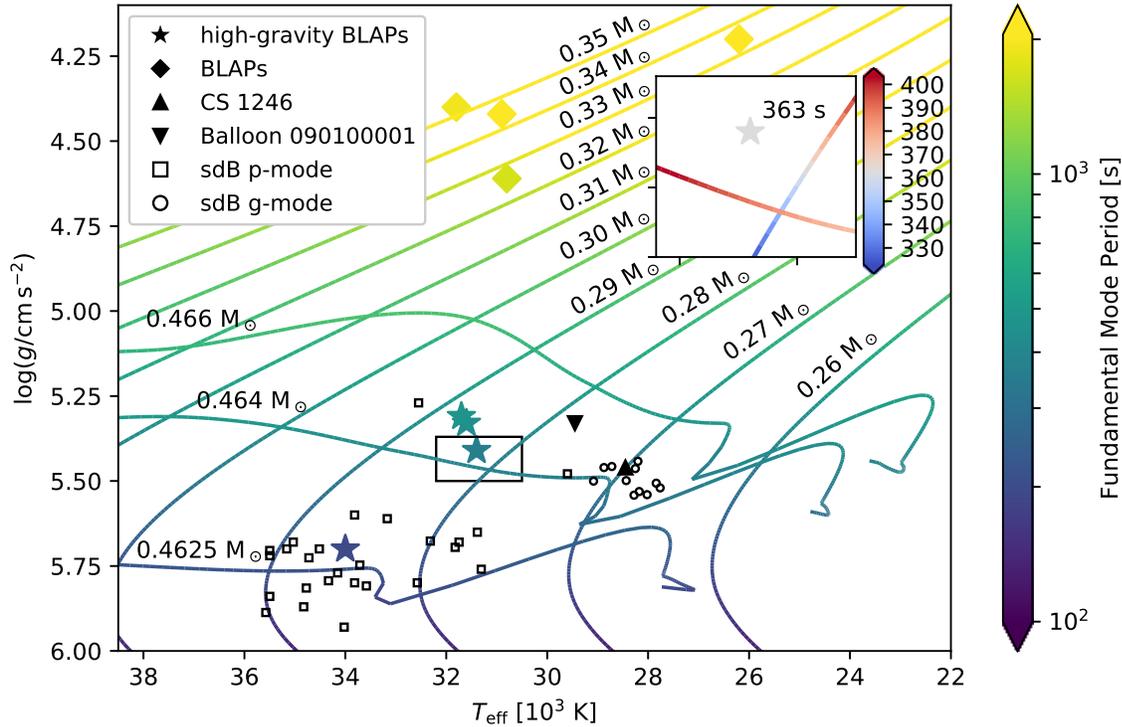}
\end{center}
\caption{\teff-\logg\, diagram with evolutionary tracks of low-mass He-core pre-WDs as well as low-mass He-burning stars with hydrogen envelopes of $0.5$, $2.0$, and $4.0 \times 10^{-3}\, M_\odot$. The color coding of the tracks corresponds to the period of the fundamental radial oscillation mode. The inset shows a zoom-in around the location of \ztfb\, bounded by the black rectangle. Overplotted are the BLAPs and our stars color coded with the observed period as well as the p-mode and g-mode sdB pulsators and the two known radial mode sdB pulsators: Balloon 090100001 and CS\,1246.}
\label{fig:teff-logg}
\end{figure} 

\subsection{Pulsation modes}
Fig.\,\ref{fig:pulsation} shows that the periodic variations in \logg\,are explained by the time-derivative of the velocity. The radius changes are a less dominant term for affecting \logg, consistent with the assumed radial mode as we show later. This implies that the atmosphere has time to adjust to the instantaneous acceleration throughout the radial pulsation cycle. Given that the sound crossing time at the photosphere ($c_s/g\approx10\,$\,sec) is noticeably less than the period, this appears likely. The large photometric amplitude, as well as the observed velocity and surface gravity shift, suggest that the observed pulsation modes are radial modes. Figure 7 in \citet{byr18} shows that unstable radial modes driven by the $\kappa$-mechanism due to the iron opacity bump are predicted for stars with our measured parameter when including radiative levitation. Our own initial non-adiabatic calculations agree with that result. We also find unstable radial modes consistent with the measured frequencies due to the iron $\kappa$-mechanism (work in preparation).

The family of hot subdwarf pulsators (including our pulsators, BLAPs, and sdB g-mode/p-mode pulsators) likely all pulsate due to $\kappa$-mechanism excitation associated with the iron opacity bump \citep{rom18}. We note that this same mechanism excites pulsation in $\beta$-Cepheids (main sequence p-mode pulsators) and slowly pulsating B-type stars (SPBs, main sequence g mode pulsators). A similar family of pulsators are those driven by the $\kappa$-mechanism of partial helium ionization, which includes Cepheids, RR Lyrae, $\delta$-Scuti, DBV stars, and pre-ELMVs (see review by \citealt{gau95}). Our high-gravity BLAPs and the previously known BLAPs are similar to high-amplitude $\delta$-Scuti and RR Lyrae stars, in the sense that they exhibit high-amplitude, low-order radial pulsation modes. The difference is that our pulsators are associated with the iron-driving instability strip and lie below the main sequence rather than above it.

\subsection{Nature of the stars}
\citet{cor18}, \citet{rom18} and \citet{byr18} proposed that the BLAPs are hot pre-helium WDs that are evolving and contracting toward the WD cooling strip, with masses approximately in the range $0.3-0.35$ \msol. To test whether the BLAPs and the new pulsators may be related, we have constructed He-core pre-WD models using the MESA stellar evolution code \citep{pax11,pax13,pax15,pax18}, release version 10398. We construct our MESA models using an initially $1.0 \, M_\odot$ star that ascends the red-giant branch (RGB), building a helium core. Once the helium core reaches a specified mass, we strip all but $0.01\, M_\odot$ of the hydrogen envelope. Residual hydrogen shell burning then governs the timescale for evolution as the star contracts and evolves toward hotter $T_{\rm eff}$ as seen in the resulting tracks in Fig.\,\ref{fig:teff-logg}. For $0.28-0.29$ \msol proto-WD models, it takes $\approx1$\,Myr to contract from \logg\ of 5.25 to 5.75. We also include adiabatic pulsation calculations using GYRE \citep{tow13} to evaluate the periods for the fundamental and first overtone radial pulsation modes at each step along these tracks.

For comparison, we also computed tracks that include period calculations from MESA models of $0.462 \, M_\odot$ low-mass He-burning stars with three different hydrogen envelope masses: $0.5$, $2.0$, and $4.0 \times 10^{-3} \, M_\odot$. As shown in Fig.\,\ref{fig:teff-logg}, all four of the new objects reside in locations where their fundamental mode periods are reasonably consistent with the observed periods for both of the scenarios. However, the inset of Fig.\,\ref{fig:teff-logg} shows a narrower region in \teff-\logg\, space near \ztfb, with a color scale centered on the measured period of $363$\,sec. The two lines in this region highlight that the expected fundamental radial mode period of a $0.28$ \msol\, He-core pre-WD is detectably different than that of a $0.464$ \msol post He-core burning star at the same \teff-\logg.

We have also calculated $f = \omega/\omega_{\rm dyn}$ for each model, with $\omega$ being the pulsation frequency, and $\omega_{\rm dyn}^2 = GM/R^3$ the stellar dynamical frequency. We find typical values of $f\approx3.6$ for the fundamental mode in low-mass He-core pre-WDs and $f\approx3.65-3.8$ and $f\approx4.95-5.2$ for the fundamental mode and the first overtone respectively in He-burning star models. Since the value of $f$ for the fundamental mode is nearly the same for He-core pre-WDs and He-burning stars, the fundamental mode periods satisfy $P_{\rm HeWD}/P_{\rm He-star} \approx \omega_{\rm dyn,He-star}/\omega_{\rm dyn,HeWD}$. At a given value of \logg, this means that $P_{\rm HeWD}/P_{\rm He-star} \approx (M_{\rm HeWD}/M_{\rm He-star})^{1/4}$, consistent with the contrast seen in the inset of Fig.\,\ref{fig:teff-logg}. This also allows us to calculate the radii and masses of the stars, assuming that $\langle$\logg$\rangle$ corresponds to the real \logg\,of each star:

\begin{equation}
R=\frac{10^{\langle\log(g)\rangle} f^2}{\omega^2} \, 
\end{equation}

\begin{equation}
M=\frac{10^{\langle\log(g)\rangle} R^2}{G} = \frac{10^{3\langle\log(g)\rangle} f^4}{G \omega^4} \, .
\end{equation}

Using these equations and $f=3.6$ we find masses of $M=0.19\pm0.05$\,\msol ($0.12\pm0.04$\,\msol; $10\times$solar), $M=0.24\pm0.07$\,\msol ($0.17\pm0.06$\,\msol; $10\times$solar), $M=0.29\pm0.08$\,\msol ($0.22\pm0.06$\,\msol; $10\times$solar) and $M=0.35\pm0.09$\,\msol ($0.27\pm0.07$\,\msol; $10\times$solar) for \ztfa, \ztfb, \ztfc\, and \ztfd\, respectively. The corresponding radii are $R=0.10\pm0.02$\,\rsol ($0.09\pm0.01$\,\rsol; $10\times$solar), $R=0.16\pm0.02$\,\rsol ($0.14\pm0.02$\,\rsol; $10\times$solar), $R=0.19\pm0.02$\,\rsol ($0.18\pm0.02$\,\rsol; $10\times$solar) and $R=0.22\pm0.03$\,\rsol ($0.20\pm0.03$\,\rsol; $10\times$solar) for \ztfa, \ztfb, \ztfc\, and \ztfd\, respectively. These masses favor the low-mass He-core pre-WDs models over the more massive post He-core burning models. 

\section{Conclusions}
We have found a new class of radial mode pulsators (high-gravity-BLAPs) with \teff\,$\approx\,30,000$\,K that are consistent with $\approx0.25-0.3$\,\msol\, low-mass He-core pre-WDs transiting the newly identified instability strip \citep{byr18,rom18} associated with the iron opacity bump when metals are enhanced from radiative levitation. With this identification, these objects are less massive analogs of the BLAPs \citep{pie17} which were also identified as He-core pre-WDs \citep{byr18,rom18}. Though these new pulsators can potentially be explained by low-mass He-shell burning stars which have evolved off the extreme horizontal branch, the predicted periods from the MESA/GYRE models provide a better match with the He-core pre-WD models. In addition, our spectroscopic discovery of the radial velocity, \logg\, and \teff\, changes associated with the high-gravity-BLAPs enables a more secure placement of these objects in the HR Diagram. More detailed asteroseismic modeling will better constrain the masses and evolutionary states of these pulsators. 

Low mass He-core pre-WDs must be formed through binary interactions (e.g. \citealt{mar95a}). High-gravity-BLAP-1 and \ztfc\, have high enough S/N in their individual spectra to measure radial velocities. In both systems, we do not find evidence for an additional radial velocity shift at ($\Delta V>25$\,\kms) after pre-whitening their pulsation-induced radial velocity changes. Therefore, we can only conclude that neither is in a compact binary with an orbital period shorter than a few hours with a $M>0.1$\,\msol\, companion. 

If the high-gravity-BLAPs are low mass He-core pre-WDs, we might be able to detect a period evolution in the pulsations ($\dot{P}$; see also \citealt{cal17}). Low-mass WDs evolve from a surface gravity of 5.25 to 5.75 and a pulsation period of $\approx500$\,sec to $\approx200$\,sec in only $\approx1$\,Myrs. This results in a substantial period drift of $\dot{P}\approx10^{-12}$\,s\,s$^{-1}$ which might be detectable after a few years monitoring. He-burning stars evolve at almost constant pulsation periods and hence are not expected to show a significant period drift, so it may be possible to confirm or rule out the He-burning evolutionary scenario by monitoring the rate of period change, even over a relatively short baseline.

Should these newly found objects instead be more massive stars that have just completed their core He-burning, we would expect an excess of pulsators at higher \logg\, near the sdB branch. If instead, both the high-gravity-BLAPs and BLAPs are He-core pre-WDs, we expect to discover more pulsators at \logg$\approx5.0$ with periods between 10-20\,min linking the two groups. This will become more apparent as the survey progresses and the sample of high-gravity-BLAPs grows.

\section{Acknowledgments}

Based on observations obtained with the Samuel Oschin Telescope 48-inch and the 60-inch Telescope at the Palomar Observatory as part of the Zwicky Transient Facility project. ZTF is supported by the National Science Foundation under Grant No. AST-1440341 and a collaboration including Caltech, IPAC, the Weizmann Institute for Science, the Oskar Klein Center at Stockholm University, the University of Maryland, the University of Washington, Deutsches Elektronen-Synchrotron and Humboldt University, Los Alamos National Laboratories, the TANGO Consortium of Taiwan, the University of Wisconsin at Milwaukee, and Lawrence Berkeley National Laboratories. Operations are conducted by COO, IPAC, and UW. This work makes use of observations from the LCOGT network.

Some of the data presented herein were obtained at the W.M. Keck Observatory, which is operated as a scientific partnership among the California Institute of Technology, the University of California and the National Aeronautics and Space Administration. The Observatory was made possible by the generous financial support of the W.M. Keck Foundation. The authors wish to recognize and acknowledge the very significant cultural role and reverence that the summit of Mauna Kea has always had within the indigenous Hawaiian community. We are most fortunate to have the opportunity to conduct observations from this mountain.

This research was supported in part by the National Science Foundation through grants AST-1514737, ACI-1663688, and at the KITP by grant PHY-1748958. This research benefited from interactions that were funded by the Gordon and Betty Moore Foundation through Grant GBMF5076. 

\vspace{5mm}
\facilities{PO:1.2m (ZTF), LCOGT (Spectral), Hale (DBSP, Chimera), Keck:I (LRIS)}

\software{\texttt{Lpipe} \citep{per19}, \texttt{PyRAF} \citep{bel16}, \texttt{Gatspy} \citep{van15}, \texttt{Period04} \citep{len05}, \texttt{FITSB2} \citep{nap04a}, \texttt{MESA} \citep{pax11,pax13,pax15,pax18}, \texttt{GYRE} \citep{tow13}, \texttt{Matplotlib} \citep{hun07}, \texttt{Astropy} \citep{astpy13, astpy18}, \texttt{Numpy} \citep{numpy}}

\bibliographystyle{aasjournal}
\bibliography{refs,refs_1508}

\begin{thebibliography}{}
\expandafter\ifx\csname natexlab\endcsname\relax\def\natexlab#1{#1}\fi
\providecommand{\url}[1]{\href{#1}{#1}}

\bibitem[{{Astropy Collaboration} {et~al.}(2013){Astropy Collaboration},
  {Robitaille}, {Tollerud}, {Greenfield}, {Droettboom}, {Bray}, {Aldcroft},
  {Davis}, {Ginsburg}, {Price-Whelan}, {Kerzendorf}, {Conley}, {Crighton},
  {Barbary}, {Muna}, {Ferguson}, {Grollier}, {Parikh}, {Nair}, {Unther},
  {Deil}, {Woillez}, {Conseil}, {Kramer}, {Turner}, {Singer}, {Fox}, {Weaver},
  {Zabalza}, {Edwards}, {Azalee Bostroem}, {Burke}, {Casey}, {Crawford},
  {Dencheva}, {Ely}, {Jenness}, {Labrie}, {Lim}, {Pierfederici}, {Pontzen},
  {Ptak}, {Refsdal}, {Servillat}, \& {Streicher}}]{astpy13}
{Astropy Collaboration}, {Robitaille}, T.~P., {Tollerud}, E.~J., {et~al.} 2013,
  \aap, 558, A33

\bibitem[{{Astropy Collaboration} {et~al.}(2018){Astropy Collaboration},
  {Price-Whelan}, {Sip{\H o}cz}, {G{\"u}nther}, {Lim}, {Crawford}, {Conseil},
  {Shupe}, {Craig}, {Dencheva}, {Ginsburg}, {VanderPlas}, {Bradley},
  {P{\'e}rez-Su{\'a}rez}, {de Val-Borro}, {Aldcroft}, {Cruz}, {Robitaille},
  {Tollerud}, {Ardelean}, {Babej}, {Bach}, {Bachetti}, {Bakanov}, {Bamford},
  {Barentsen}, {Barmby}, {Baumbach}, {Berry}, {Biscani}, {Boquien}, {Bostroem},
  {Bouma}, {Brammer}, {Bray}, {Breytenbach}, {Buddelmeijer}, {Burke},
  {Calderone}, {Cano Rodr{\'{\i}}guez}, {Cara}, {Cardoso}, {Cheedella},
  {Copin}, {Corrales}, {Crichton}, {D'Avella}, {Deil}, {Depagne}, {Dietrich},
  {Donath}, {Droettboom}, {Earl}, {Erben}, {Fabbro}, {Ferreira}, {Finethy},
  {Fox}, {Garrison}, {Gibbons}, {Goldstein}, {Gommers}, {Greco}, {Greenfield},
  {Groener}, {Grollier}, {Hagen}, {Hirst}, {Homeier}, {Horton}, {Hosseinzadeh},
  {Hu}, {Hunkeler}, {Ivezi{\'c}}, {Jain}, {Jenness}, {Kanarek}, {Kendrew},
  {Kern}, {Kerzendorf}, {Khvalko}, {King}, {Kirkby}, {Kulkarni}, {Kumar},
  {Lee}, {Lenz}, {Littlefair}, {Ma}, {Macleod}, {Mastropietro}, {McCully},
  {Montagnac}, {Morris}, {Mueller}, {Mumford}, {Muna}, {Murphy}, {Nelson},
  {Nguyen}, {Ninan}, {N{\"o}the}, {Ogaz}, {Oh}, {Parejko}, {Parley}, {Pascual},
  {Patil}, {Patil}, {Plunkett}, {Prochaska}, {Rastogi}, {Reddy Janga},
  {Sabater}, {Sakurikar}, {Seifert}, {Sherbert}, {Sherwood-Taylor}, {Shih},
  {Sick}, {Silbiger}, {Singanamalla}, {Singer}, {Sladen}, {Sooley},
  {Sornarajah}, {Streicher}, {Teuben}, {Thomas}, {Tremblay}, {Turner},
  {Terr{\'o}n}, {van Kerkwijk}, {de la Vega}, {Watkins}, {Weaver}, {Whitmore},
  {Woillez}, {Zabalza}, \& {Astropy Contributors}}]{astpy18}
{Astropy Collaboration}, {Price-Whelan}, A.~M., {Sip{\H o}cz}, B.~M., {et~al.}
  2018, \aj, 156, 123

\bibitem[{{Bailer-Jones} {et~al.}(2018){Bailer-Jones}, {Rybizki}, {Fouesneau},
  {Mantelet}, \& {Andrae}}]{bai18}
{Bailer-Jones}, C.~A.~L., {Rybizki}, J., {Fouesneau}, M., {Mantelet}, G., \&
  {Andrae}, R. 2018, \aj, 156, 58

\bibitem[{{Barlow} {et~al.}(2010){Barlow}, {Dunlap}, {Clemens}, {Lynas-Gray},
  {Ivarsen}, {Lacluyze}, {Reichart}, {Haislip}, \& {Nysewander}}]{bar10}
{Barlow}, B.~N., {Dunlap}, B.~H., {Clemens}, J.~C., {et~al.} 2010, \mnras, 403,
  324

\bibitem[{{Bellm} \& {Sesar}(2016)}]{bel16}
{Bellm}, E.~C., \& {Sesar}, B. 2016, {pyraf-dbsp: Reduction pipeline for the
  Palomar Double Beam Spectrograph}, Astrophysics Source Code Library, , ,
  ascl:1602.002

\bibitem[{{Bellm} {et~al.}(2019{\natexlab{a}}){Bellm}, {Kulkarni}, {Graham},
  {Dekany}, {Smith}, {Riddle}, {Masci}, {Helou}, {Prince}, {Adams},
  {Barbarino}, {Barlow}, {Bauer}, {Beck}, {Belicki}, {Biswas}, {Blagorodnova},
  {Bodewits}, {Bolin}, {Brinnel}, {Brooke}, {Bue}, {Bulla}, {Burruss}, {Cenko},
  {Chang}, {Connolly}, {Coughlin}, {Cromer}, {Cunningham}, {De}, {Delacroix},
  {Desai}, {Duev}, {Eadie}, {Farnham}, {Feeney}, {Feindt}, {Flynn},
  {Franckowiak}, {Frederick}, {Fremling}, {Gal-Yam}, {Gezari}, {Giomi},
  {Goldstein}, {Golkhou}, {Goobar}, {Groom}, {Hacopians}, {Hale}, {Henning},
  {Ho}, {Hover}, {Howell}, {Hung}, {Huppenkothen}, {Imel}, {Ip}, {Ivezi{\'c}},
  {Jackson}, {Jones}, {Juric}, {Kasliwal}, {Kaspi}, {Kaye}, {Kelley},
  {Kowalski}, {Kramer}, {Kupfer}, {Landry}, {Laher}, {Lee}, {Lin}, {Lin},
  {Lunnan}, {Giomi}, {Mahabal}, {Mao}, {Miller}, {Monkewitz}, {Murphy},
  {Ngeow}, {Nordin}, {Nugent}, {Ofek}, {Patterson}, {Penprase}, {Porter},
  {Rauch}, {Rebbapragada}, {Reiley}, {Rigault}, {Rodriguez}, {van Roestel},
  {Rusholme}, {van Santen}, {Schulze}, {Shupe}, {Singer}, {Soumagnac}, {Stein},
  {Surace}, {Sollerman}, {Szkody}, {Taddia}, {Terek}, {Van Sistine}, {van
  Velzen}, {Vestrand}, {Walters}, {Ward}, {Ye}, {Yu}, {Yan}, \&
  {Zolkower}}]{bel19}
{Bellm}, E.~C., {Kulkarni}, S.~R., {Graham}, M.~J., {et~al.}
  2019{\natexlab{a}}, \pasp, 131, 018002

\bibitem[{{Bellm} {et~al.}(2019{\natexlab{b}}){Bellm}, {Kulkarni}, {Barlow},
  {Feindt}, {Graham}, {Goobar}, {Kupfer}, {Ngeow}, {Nugent}, {Ofek}, {Prince},
  {Riddle}, {Walters}, \& {Ye}}]{bel19a}
{Bellm}, E.~C., {Kulkarni}, S.~R., {Barlow}, T., {et~al.} 2019{\natexlab{b}},
  \pasp, 131, 068003

\bibitem[{{Brown} {et~al.}(2013){Brown}, {Baliber}, {Bianco}, {Bowman},
  {Burleson}, {Conway}, {Crellin}, {Depagne}, {De Vera}, {Dilday}, {Dragomir},
  {Dubberley}, {Eastman}, {Elphick}, {Falarski}, {Foale}, {Ford}, {Fulton},
  {Garza}, {Gomez}, {Graham}, {Greene}, {Haldeman}, {Hawkins}, {Haworth},
  {Haynes}, {Hidas}, {Hjelstrom}, {Howell}, {Hygelund}, {Lister}, {Lobdill},
  {Martinez}, {Mullins}, {Norbury}, {Parrent}, {Paulson}, {Petry}, {Pickles},
  {Posner}, {Rosing}, {Ross}, {Sand}, {Saunders}, {Shobbrook}, {Shporer},
  {Street}, {Thomas}, {Tsapras}, {Tufts}, {Valenti}, {Vander Horst}, {Walker},
  {White}, \& {Willis}}]{bro13}
{Brown}, T.~M., {Baliber}, N., {Bianco}, F.~B., {et~al.} 2013, \pasp, 125, 1031

\bibitem[{{Byrne} \& {Jeffery}(2018)}]{byr18}
{Byrne}, C.~M., \& {Jeffery}, C.~S. 2018, \mnras, 481, 3810

\bibitem[{{Calcaferro} {et~al.}(2017){Calcaferro}, {C{\'o}rsico}, \&
  {Althaus}}]{cal17}
{Calcaferro}, L.~M., {C{\'o}rsico}, A.~H., \& {Althaus}, L.~G. 2017, \aap, 600,
  A73

\bibitem[{{Chambers} {et~al.}(2016){Chambers}, {Magnier}, {Metcalfe},
  {Flewelling}, {Huber}, {Waters}, {Denneau}, {Draper}, {Farrow}, {Finkbeiner},
  {Holmberg}, {Koppenhoefer}, {Price}, {Saglia}, {Schlafly}, {Smartt},
  {Sweeney}, {Wainscoat}, {Burgett}, {Grav}, {Heasley}, {Hodapp}, {Jedicke},
  {Kaiser}, {Kudritzki}, {Luppino}, {Lupton}, {Monet}, {Morgan}, {Onaka},
  {Stubbs}, {Tonry}, {Banados}, {Bell}, {Bender}, {Bernard}, {Botticella},
  {Casertano}, {Chastel}, {Chen}, {Chen}, {Cole}, {Deacon}, {Frenk},
  {Fitzsimmons}, {Gezari}, {Goessl}, {Goggia}, {Goldman}, {Grebel}, {Hambly},
  {Hasinger}, {Heavens}, {Heckman}, {Henderson}, {Henning}, {Holman}, {Hopp},
  {Ip}, {Isani}, {Keyes}, {Koekemoer}, {Kotak}, {Long}, {Lucey}, {Liu},
  {Martin}, {McLean}, {Morganson}, {Murphy}, {Nieto-Santisteban}, {Norberg},
  {Peacock}, {Pier}, {Postman}, {Primak}, {Rae}, {Rest}, {Riess}, {Riffeser},
  {Rix}, {Roser}, {Schilbach}, {Schultz}, {Scolnic}, {Szalay}, {Seitz},
  {Shiao}, {Small}, {Smith}, {Soderblom}, {Taylor}, {Thakar}, {Thiel},
  {Thilker}, {Urata}, {Valenti}, {Walter}, {Watters}, {Werner}, {White},
  {Wood-Vasey}, \& {Wyse}}]{cham16}
{Chambers}, K.~C., {Magnier}, E.~A., {Metcalfe}, N., {et~al.} 2016, arXiv
  e-prints, arXiv:1612.05560

\bibitem[{{Charpinet} {et~al.}(1997){Charpinet}, {Fontaine}, {Brassard},
  {Chayer}, {Rogers}, {Iglesias}, \& {Dorman}}]{cha97}
{Charpinet}, S., {Fontaine}, G., {Brassard}, P., {et~al.} 1997, \apjl, 483,
  L123

\bibitem[{{Charpinet} {et~al.}(1996){Charpinet}, {Fontaine}, {Brassard}, \&
  {Dorman}}]{cha96}
{Charpinet}, S., {Fontaine}, G., {Brassard}, P., \& {Dorman}, B. 1996, \apjl,
  471, L103

\bibitem[{{C{\'o}rsico} {et~al.}(2016){C{\'o}rsico}, {Althaus}, {Serenelli},
  {Kepler}, {Jeffery}, \& {Corti}}]{cor16}
{C{\'o}rsico}, A.~H., {Althaus}, L.~G., {Serenelli}, A.~M., {et~al.} 2016,
  \aap, 588, A74

\bibitem[{{C{\'o}rsico} {et~al.}(2018){C{\'o}rsico}, {Romero}, {Althaus},
  {Pelisoli}, \& {Kepler}}]{cor18}
{C{\'o}rsico}, A.~H., {Romero}, A.~D., {Althaus}, L.~G., {Pelisoli}, I., \&
  {Kepler}, S.~O. 2018, arXiv e-prints, arXiv:1809.07451

\bibitem[{{Fontaine} {et~al.}(2003){Fontaine}, {Brassard}, {Charpinet},
  {Green}, {Chayer}, {Bill{\`e}res}, \& {Randall}}]{fon03}
{Fontaine}, G., {Brassard}, P., {Charpinet}, S., {et~al.} 2003, \apj, 597, 518

\bibitem[{{Gaia Collaboration} {et~al.}(2018){Gaia Collaboration}, {Brown},
  {Vallenari}, {Prusti}, {de Bruijne}, {Babusiaux}, {Bailer-Jones}, {Biermann},
  {Evans}, {Eyer}, {Jansen}, {Jordi}, {Klioner}, {Lammers}, {Lindegren},
  {Luri}, {Mignard}, {Panem}, {Pourbaix}, {Randich}, {Sartoretti}, {Siddiqui},
  {Soubiran}, {van Leeuwen}, {Walton}, {Arenou}, {Bastian}, {Cropper},
  {Drimmel}, {Katz}, {Lattanzi}, {Bakker}, {Cacciari}, {Casta{\~n}eda},
  {Chaoul}, {Cheek}, {De Angeli}, {Fabricius}, {Guerra}, {Holl}, {Masana},
  {Messineo}, {Mowlavi}, {Nienartowicz}, {Panuzzo}, {Portell}, {Riello},
  {Seabroke}, {Tanga}, {Th{\'e}venin}, {Gracia-Abril}, {Comoretto},
  {Garcia-Reinaldos}, {Teyssier}, {Altmann}, {Andrae}, {Audard},
  {Bellas-Velidis}, {Benson}, {Berthier}, {Blomme}, {Burgess}, {Busso},
  {Carry}, {Cellino}, {Clementini}, {Clotet}, {Creevey}, {Davidson}, {De
  Ridder}, {Delchambre}, {Dell'Oro}, {Ducourant},
  {Fern{\'a}ndez-Hern{\'a}ndez}, {Fouesneau}, {Fr{\'e}mat}, {Galluccio},
  {Garc{\'\i}a-Torres}, {Gonz{\'a}lez-N{\'u}{\~n}ez}, {Gonz{\'a}lez-Vidal},
  {Gosset}, {Guy}, {Halbwachs}, {Hambly}, {Harrison}, {Hern{\'a}ndez},
  {Hestroffer}, {Hodgkin}, {Hutton}, {Jasniewicz}, {Jean-Antoine-Piccolo},
  {Jordan}, {Korn}, {Krone-Martins}, {Lanzafame}, {Lebzelter}, {L{\"o}ffler},
  {Manteiga}, {Marrese}, {Mart{\'\i}n-Fleitas}, {Moitinho}, {Mora}, {Muinonen},
  {Osinde}, {Pancino}, {Pauwels}, {Petit}, {Recio-Blanco}, {Richards},
  {Rimoldini}, {Robin}, {Sarro}, {Siopis}, {Smith}, {Sozzetti}, {S{\"u}veges},
  {Torra}, {van Reeven}, {Abbas}, {Abreu Aramburu}, {Accart}, {Aerts},
  {Altavilla}, {{\'A}lvarez}, {Alvarez}, {Alves}, {Anderson}, {Andrei},
  {Anglada Varela}, {Antiche}, {Antoja}, {Arcay}, {Astraatmadja}, {Bach},
  {Baker}, {Balaguer-N{\'u}{\~n}ez}, {Balm}, {Barache}, {Barata}, {Barbato},
  {Barblan}, {Barklem}, {Barrado}, {Barros}, {Barstow}, {Bartholom{\'e}
  Mu{\~n}oz}, {Bassilana}, {Becciani}, {Bellazzini}, {Berihuete}, {Bertone},
  {Bianchi}, {Bienaym{\'e}}, {Blanco-Cuaresma}, {Boch}, {Boeche}, {Bombrun},
  {Borrachero}, {Bossini}, {Bouquillon}, {Bourda}, {Bragaglia}, {Bramante},
  {Breddels}, {Bressan}, {Brouillet}, {Br{\"u}semeister}, {Brugaletta},
  {Bucciarelli}, {Burlacu}, {Busonero}, {Butkevich}, {Buzzi}, {Caffau},
  {Cancelliere}, {Cannizzaro}, {Cantat-Gaudin}, {Carballo}, {Carlucci},
  {Carrasco}, {Casamiquela}, {Castellani}, {Castro-Ginard}, {Charlot},
  {Chemin}, {Chiavassa}, {Cocozza}, {Costigan}, {Cowell}, {Crifo}, {Crosta},
  {Crowley}, {Cuypers}, {Dafonte}, {Damerdji}, {Dapergolas}, {David}, {David},
  {de Laverny}, {De Luise}, {De March}, {de Martino}, {de Souza}, {de Torres},
  {Debosscher}, {del Pozo}, {Delbo}, {Delgado}, {Delgado}, {Di Matteo},
  {Diakite}, {Diener}, {Distefano}, {Dolding}, {Drazinos}, {Dur{\'a}n},
  {Edvardsson}, {Enke}, {Eriksson}, {Esquej}, {Eynard Bontemps}, {Fabre},
  {Fabrizio}, {Faigler}, {Falc{\~a}o}, {Farr{\`a}s Casas}, {Federici},
  {Fedorets}, {Fernique}, {Figueras}, {Filippi}, {Findeisen}, {Fonti},
  {Fraile}, {Fraser}, {Fr{\'e}zouls}, {Gai}, {Galleti}, {Garabato},
  {Garc{\'\i}a-Sedano}, {Garofalo}, {Garralda}, {Gavel}, {Gavras}, {Gerssen},
  {Geyer}, {Giacobbe}, {Gilmore}, {Girona}, {Giuffrida}, {Glass}, {Gomes},
  {Granvik}, {Gueguen}, {Guerrier}, {Guiraud}, {Guti{\'e}rrez-S{\'a}nchez},
  {Haigron}, {Hatzidimitriou}, {Hauser}, {Haywood}, {Heiter}, {Helmi}, {Heu},
  {Hilger}, {Hobbs}, {Hofmann}, {Holland}, {Huckle}, {Hypki}, {Icardi},
  {Jan{\ss}en}, {Jevardat de Fombelle}, {Jonker}, {Juh{\'a}sz}, {Julbe},
  {Karampelas}, {Kewley}, {Klar}, {Kochoska}, {Kohley}, {Kolenberg},
  {Kontizas}, {Kontizas}, {Koposov}, {Kordopatis}, {Kostrzewa-Rutkowska},
  {Koubsky}, {Lambert}, {Lanza}, {Lasne}, {Lavigne}, {Le Fustec}, {Le
  Poncin-Lafitte}, {Lebreton}, {Leccia}, {Leclerc}, {Lecoeur-Taibi},
  {Lenhardt}, {Leroux}, {Liao}, {Licata}, {Lindstr{\o}m}, {Lister}, {Livanou},
  {Lobel}, {L{\'o}pez}, {Managau}, {Mann}, {Mantelet}, {Marchal}, {Marchant},
  {Marconi}, {Marinoni}, {Marschalk{\'o}}, {Marshall}, {Martino}, {Marton},
  {Mary}, {Massari}, {Matijevi{\v{c}}}, {Mazeh}, {McMillan}, {Messina},
  {Michalik}, {Millar}, {Molina}, {Molinaro}, {Moln{\'a}r}, {Montegriffo},
  {Mor}, {Morbidelli}, {Morel}, {Morris}, {Mulone}, {Muraveva}, {Musella},
  {Nelemans}, {Nicastro}, {Noval}, {O'Mullane}, {Ord{\'e}novic},
  {Ord{\'o}{\~n}ez-Blanco}, {Osborne}, {Pagani}, {Pagano}, {Pailler},
  {Palacin}, {Palaversa}, {Panahi}, {Pawlak}, {Piersimoni}, {Pineau}, {Plachy},
  {Plum}, {Poggio}, {Poujoulet}, {Pr{\v{s}}a}, {Pulone}, {Racero}, {Ragaini},
  {Rambaux}, {Ramos-Lerate}, {Regibo}, {Reyl{\'e}}, {Riclet}, {Ripepi}, {Riva},
  {Rivard}, {Rixon}, {Roegiers}, {Roelens}, {Romero-G{\'o}mez}, {Rowell},
  {Royer}, {Ruiz-Dern}, {Sadowski}, {Sagrist{\`a} Sell{\'e}s}, {Sahlmann},
  {Salgado}, {Salguero}, {Sanna}, {Santana-Ros}, {Sarasso}, {Savietto},
  {Schultheis}, {Sciacca}, {Segol}, {Segovia}, {S{\'e}gransan}, {Shih},
  {Siltala}, {Silva}, {Smart}, {Smith}, {Solano}, {Solitro}, {Sordo}, {Soria
  Nieto}, {Souchay}, {Spagna}, {Spoto}, {Stampa}, {Steele},
  {Steidelm{\"u}ller}, {Stephenson}, {Stoev}, {Suess}, {Surdej}, {Szabados},
  {Szegedi-Elek}, {Tapiador}, {Taris}, {Tauran}, {Taylor}, {Teixeira},
  {Terrett}, {Teyssand ier}, {Thuillot}, {Titarenko}, {Torra Clotet}, {Turon},
  {Ulla}, {Utrilla}, {Uzzi}, {Vaillant}, {Valentini}, {Valette}, {van Elteren},
  {Van Hemelryck}, {van Leeuwen}, {Vaschetto}, {Vecchiato}, {Veljanoski},
  {Viala}, {Vicente}, {Vogt}, {von Essen}, {Voss}, {Votruba}, {Voutsinas},
  {Walmsley}, {Weiler}, {Wertz}, {Wevers}, {Wyrzykowski}, {Yoldas},
  {{\v{Z}}erjal}, {Ziaeepour}, {Zorec}, {Zschocke}, {Zucker}, {Zurbach}, \&
  {Zwitter}}]{gai18}
{Gaia Collaboration}, {Brown}, A.~G.~A., {Vallenari}, A., {et~al.} 2018, \aap,
  616, A1

\bibitem[{{Gautschy} \& {Saio}(1995)}]{gau95}
{Gautschy}, A., \& {Saio}, H. 1995, \araa, 33, 75

\bibitem[{{Geier} {et~al.}(2019){Geier}, {Raddi}, {Gentile Fusillo}, \&
  {Marsh}}]{gei19}
{Geier}, S., {Raddi}, R., {Gentile Fusillo}, N.~P., \& {Marsh}, T.~R. 2019,
  \aap, 621, A38

\bibitem[{{Gianninas} {et~al.}(2016){Gianninas}, {Curd}, {Fontaine}, {Brown},
  \& {Kilic}}]{gia16}
{Gianninas}, A., {Curd}, B., {Fontaine}, G., {Brown}, W.~R., \& {Kilic}, M.
  2016, \apjl, 822, L27

\bibitem[{{Graham} {et~al.}(2013){Graham}, {Drake}, {Djorgovski}, {Mahabal}, \&
  {Donalek}}]{gra13}
{Graham}, M.~J., {Drake}, A.~J., {Djorgovski}, S.~G., {Mahabal}, A.~A., \&
  {Donalek}, C. 2013, \mnras, 434, 2629

\bibitem[{{Graham} {et~al.}(2019){Graham}, {Kulkarni}, {Bellm}, {Adams},
  {Barbarino}, {Blagorodnova}, {Bodewits}, {Bolin}, {Brady}, {Cenko}, {Chang},
  {Coughlin}, {De}, {Eadie}, {Farnham}, {Feindt}, {Franckowiak}, {Fremling},
  {Gezari}, {Ghosh}, {Goldstein}, {Golkhou}, {Goobar}, {Ho}, {Huppenkothen},
  {Ivezi{\'c}}, {Jones}, {Juric}, {Kaplan}, {Kasliwal}, {Kelley}, {Kupfer},
  {Lee}, {Lin}, {Lunnan}, {Mahabal}, {Miller}, {Ngeow}, {Nugent}, {Ofek},
  {Prince}, {Rauch}, {van Roestel}, {Schulze}, {Singer}, {Sollerman}, {Taddia},
  {Yan}, {Ye}, {Yu}, {Barlow}, {Bauer}, {Beck}, {Belicki}, {Biswas}, {Brinnel},
  {Brooke}, {Bue}, {Bulla}, {Burruss}, {Connolly}, {Cromer}, {Cunningham},
  {Dekany}, {Delacroix}, {Desai}, {Duev}, {Feeney}, {Flynn}, {Frederick},
  {Gal-Yam}, {Giomi}, {Groom}, {Hacopians}, {Hale}, {Helou}, {Henning},
  {Hover}, {Hillenbrand}, {Howell}, {Hung}, {Imel}, {Ip}, {Jackson}, {Kaspi},
  {Kaye}, {Kowalski}, {Kramer}, {Kuhn}, {Landry}, {Laher}, {Mao}, {Masci},
  {Monkewitz}, {Murphy}, {Nordin}, {Patterson}, {Penprase}, {Porter},
  {Rebbapragada}, {Reiley}, {Riddle}, {Rigault}, {Rodriguez}, {Rusholme}, {van
  Santen}, {Shupe}, {Smith}, {Soumagnac}, {Stein}, {Surace}, {Szkody}, {Terek},
  {Van Sistine}, {van Velzen}, {Vestrand}, {Walters}, {Ward}, {Zhang}, \&
  {Zolkower}}]{gra19}
{Graham}, M.~J., {Kulkarni}, S.~R., {Bellm}, E.~C., {et~al.} 2019, \pasp, 131,
  078001

\bibitem[{{Green} {et~al.}(2003){Green}, {Callerame}, {Seitenzahl}, {White},
  {Hyde}, {Giovanni}, {Reed}, {Fontaine}, \& {{\O}stensen}}]{gre03}
{Green}, E.~M., {Callerame}, K., {Seitenzahl}, I.~R., {et~al.} 2003, \apss,
  284, 65

\bibitem[{{Green} {et~al.}(2018){Green}, {Schlafly}, {Finkbeiner}, {Rix},
  {Martin}, {Burgett}, {Draper}, {Flewelling}, {Hodapp}, {Kaiser}, {Kudritzki},
  {Magnier}, {Metcalfe}, {Tonry}, {Wainscoat}, \& {Waters}}]{gre18}
{Green}, G.~M., {Schlafly}, E.~F., {Finkbeiner}, D., {et~al.} 2018, \mnras,
  478, 651

\bibitem[{{Harding} {et~al.}(2016){Harding}, {Hallinan}, {Milburn}, {Gardner},
  {Konidaris}, {Singh}, {Shao}, {Sandhu}, {Kyne}, \& {Schlichting}}]{har16}
{Harding}, L.~K., {Hallinan}, G., {Milburn}, J., {et~al.} 2016, \mnras, 457,
  3036

\bibitem[{{Heber}(1986)}]{heb86}
{Heber}, U. 1986, \aap, 155, 33

\bibitem[{{Heber}(2009)}]{heb09}
---. 2009, \araa, 47, 211

\bibitem[{{Heber}(2016)}]{heb16}
---. 2016, \pasp, 128, 082001

\bibitem[{{Heber} {et~al.}(2000){Heber}, {Reid}, \& {Werner}}]{heb00}
{Heber}, U., {Reid}, I.~N., \& {Werner}, K. 2000, \aap, 363, 198

\bibitem[{Hunter(2007)}]{hun07}
Hunter, J.~D. 2007, Computing In Science \& Engineering, 9, 90

\bibitem[{{Istrate} {et~al.}(2017){Istrate}, {Fontaine}, \& {Heuser}}]{ist17}
{Istrate}, A.~G., {Fontaine}, G., \& {Heuser}, C. 2017, \apj, 847, 130

\bibitem[{{Jeffery} \& {Saio}(2013)}]{jef13}
{Jeffery}, C.~S., \& {Saio}, H. 2013, \mnras, 435, 885

\bibitem[{{Kilkenny} {et~al.}(1997){Kilkenny}, {Koen}, {O'Donoghue}, \&
  {Stobie}}]{kil97}
{Kilkenny}, D., {Koen}, C., {O'Donoghue}, D., \& {Stobie}, R.~S. 1997, \mnras,
  285, 640

\bibitem[{{Lenz} \& {Breger}(2005)}]{len05}
{Lenz}, P., \& {Breger}, M. 2005, Communications in Asteroseismology, 146, 53

\bibitem[{{Lomb}(1976)}]{Lom76}
{Lomb}, N.~R. 1976, \apss, 39, 447

\bibitem[{{Marsh} {et~al.}(1995){Marsh}, {Dhillon}, \& {Duck}}]{mar95a}
{Marsh}, T.~R., {Dhillon}, V.~S., \& {Duck}, S.~R. 1995, \mnras, 275, 828

\bibitem[{{Masci} {et~al.}(2019){Masci}, {Laher}, {Rusholme}, {Shupe}, {Groom},
  {Surace}, {Jackson}, {Monkewitz}, {Beck}, {Flynn}, {Terek}, {Landry},
  {Hacopians}, {Desai}, {Howell}, {Brooke}, {Imel}, {Wachter}, {Ye}, {Lin},
  {Cenko}, {Cunningham}, {Rebbapragada}, {Bue}, {Miller}, {Mahabal}, {Bellm},
  {Patterson}, {Juri{\'c}}, {Golkhou}, {Ofek}, {Walters}, {Graham}, {Kasliwal},
  {Dekany}, {Kupfer}, {Burdge}, {Cannella}, {Barlow}, {Van Sistine}, {Giomi},
  {Fremling}, {Blagorodnova}, {Levitan}, {Riddle}, {Smith}, {Helou}, {Prince},
  \& {Kulkarni}}]{mas19}
{Masci}, F.~J., {Laher}, R.~R., {Rusholme}, B., {et~al.} 2019, \pasp, 131,
  018003

\bibitem[{{Maxted} {et~al.}(2014){Maxted}, {Serenelli}, {Marsh}, {Catal{\'a}n},
  {Mahtani}, \& {Dhillon}}]{max14}
{Maxted}, P.~F.~L., {Serenelli}, A.~M., {Marsh}, T.~R., {et~al.} 2014, \mnras,
  444, 208

\bibitem[{{Maxted} {et~al.}(2013){Maxted}, {Serenelli}, {Miglio}, {Marsh},
  {Heber}, {Dhillon}, {Littlefair}, {Copperwheat}, {Smalley}, {Breedt}, \&
  {Schaffenroth}}]{max13}
{Maxted}, P.~F.~L., {Serenelli}, A.~M., {Miglio}, A., {et~al.} 2013, \nat, 498,
  463

\bibitem[{{McCarthy} {et~al.}(1998){McCarthy}, {Cohen}, {Butcher}, {Cromer},
  {Croner}, {Douglas}, {Goeden}, {Grewal}, {Lu}, {Petrie}, {Weng}, {Weber},
  {Koch}, \& {Rodgers}}]{mcc98}
{McCarthy}, J.~K., {Cohen}, J.~G., {Butcher}, B., {et~al.} 1998, in \procspie,
  Vol. 3355, Optical Astronomical Instrumentation, ed. S.~{D'Odorico}, 81--92

\bibitem[{{Napiwotzki} {et~al.}(2004){Napiwotzki}, {Karl}, {Lisker}, {Heber},
  {Christlieb}, {Reimers}, {Nelemans}, \& {Homeier}}]{nap04a}
{Napiwotzki}, R., {Karl}, C.~A., {Lisker}, T., {et~al.} 2004, Astrophysics and
  Space Science, 291, 321

\bibitem[{{Oke} \& {Gunn}(1982)}]{oke82}
{Oke}, J.~B., \& {Gunn}, J.~E. 1982, PASP, 94, 586

\bibitem[{Oliphant(2015)}]{numpy}
Oliphant, T.~E. 2015, Guide to NumPy, 2nd edn. (USA: CreateSpace Independent
  Publishing Platform)

\bibitem[{{Oreiro} {et~al.}(2004){Oreiro}, {Ulla}, {P{\'e}rez Hern{\'a}ndez},
  {{\O}stensen}, {Rodr{\'\i}guez L{\'o}pez}, \& {MacDonald}}]{ore04}
{Oreiro}, R., {Ulla}, A., {P{\'e}rez Hern{\'a}ndez}, F., {et~al.} 2004, \aap,
  418, 243

\bibitem[{{{\O}stensen} {et~al.}(2010){{\O}stensen}, {Silvotti}, {Charpinet},
  {Oreiro}, {Handler}, {Green}, {Bloemen}, {Heber}, {G{\"a}nsicke}, {Marsh},
  {Kurtz}, {Telting}, {Reed}, {Kawaler}, {Aerts}, {Rodr{\'\i}guez-L{\'o}pez},
  {Vu{\v{c}}kovi{\'c}}, {Ottosen}, {Liimets}, {Quint}, {Van Grootel},
  {Randall}, {Gilliland }, {Kjeldsen}, {Christensen-Dalsgaard}, {Borucki},
  {Koch}, \& {Quintana}}]{ost10}
{{\O}stensen}, R.~H., {Silvotti}, R., {Charpinet}, S., {et~al.} 2010, \mnras,
  409, 1470

\bibitem[{{O'Toole} \& {Heber}(2006)}]{oto06}
{O'Toole}, S.~J., \& {Heber}, U. 2006, \aap, 452, 579

\bibitem[{{Paxton} {et~al.}(2011){Paxton}, {Bildsten}, {Dotter}, {Herwig},
  {Lesaffre}, \& {Timmes}}]{pax11}
{Paxton}, B., {Bildsten}, L., {Dotter}, A., {et~al.} 2011, ApJs, 192, 3

\bibitem[{{Paxton} {et~al.}(2013){Paxton}, {Cantiello}, {Arras}, {Bildsten},
  {Brown}, {Dotter}, {Mankovich}, {Montgomery}, {Stello}, {Timmes}, \&
  {Townsend}}]{pax13}
{Paxton}, B., {Cantiello}, M., {Arras}, P., {et~al.} 2013, ApJs, 208, 4

\bibitem[{{Paxton} {et~al.}(2015){Paxton}, {Marchant}, {Schwab}, {Bauer},
  {Bildsten}, {Cantiello}, {Dessart}, {Farmer}, {Hu}, {Langer}, {Townsend},
  {Townsley}, \& {Timmes}}]{pax15}
{Paxton}, B., {Marchant}, P., {Schwab}, J., {et~al.} 2015, ApJs, 220, 15

\bibitem[{{Paxton} {et~al.}(2018){Paxton}, {Schwab}, {Bauer}, {Bildsten},
  {Blinnikov}, {Duffell}, {Farmer}, {Goldberg}, {Marchant}, {Sorokina},
  {Thoul}, {Townsend}, \& {Timmes}}]{pax18}
{Paxton}, B., {Schwab}, J., {Bauer}, E.~B., {et~al.} 2018, \apjs, 234, 34

\bibitem[{{Perley}(2019)}]{per19}
{Perley}, D.~A. 2019, arXiv e-prints, arXiv:1903.07629

\bibitem[{{Pietrukowicz} {et~al.}(2017){Pietrukowicz}, {Dziembowski}, {Latour},
  {Angeloni}, {Poleski}, {di Mille}, {Soszy{\'n}ski}, {Udalski},
  {Szyma{\'n}ski}, {Wyrzykowski}, {Koz{\l}owski}, {Skowron}, {Skowron},
  {Mr{\'o}z}, {Pawlak}, \& {Ulaczyk}}]{pie17}
{Pietrukowicz}, P., {Dziembowski}, W.~A., {Latour}, M., {et~al.} 2017, Nature
  Astronomy, 1, 0166

\bibitem[{{Romero} {et~al.}(2018){Romero}, {C{\'o}rsico}, {Althaus},
  {Pelisoli}, \& {Kepler}}]{rom18}
{Romero}, A.~D., {C{\'o}rsico}, A.~H., {Althaus}, L.~G., {Pelisoli}, I., \&
  {Kepler}, S.~O. 2018, \mnras, 477, L30

\bibitem[{{Scargle}(1982)}]{sca82}
{Scargle}, J.~D. 1982, \apj, 263, 835

\bibitem[{{Townsend} \& {Teitler}(2013)}]{tow13}
{Townsend}, R.~H.~D., \& {Teitler}, S.~A. 2013, \mnras, 435, 3406

\bibitem[{{VanderPlas} \& {Ivezi\'{c}}(2015)}]{van15}
{VanderPlas}, J.~T., \& {Ivezi\'{c}}, v. 2015, \apj, 812, 18

\end{thebibliography}

\end{document}